\documentclass[aps,pre,twocolumn,showpacs,superscriptaddress,groupedaddress,10pt]{revtex4}
\renewcommand \thesection{\Roman{section}}
\usepackage{graphicx}
\usepackage{color} 
\usepackage{epstopdf}
\usepackage{epsfig}
\usepackage{amssymb}
\usepackage{amsmath}
\usepackage{setspace}
\usepackage{float}

\begin{document}

\title{Interplay between membrane elasticity and active cytoskeleton
  forces regulates the aggregation dynamics of the immunological
  synapse}

\author{Nadiv Dharan} \affiliation{Department of Biomedical
  Engineering, Ben Gurion University of the Negev, Be'er Sheva 84105,
  Israel}

\author{Oded Farago} \affiliation{Department of Biomedical
  Engineering, Ben Gurion University of the Negev, Be'er Sheva 84105,
  Israel} \affiliation{Ilse Katz Institute for Nanoscale Science and
  Technology, Ben Gurion University of the Negev, Be'er Sheva 84105,
  Israel}

\begin{abstract}

Adhesion between a T cell and an antigen presenting cell is achieved
by TCR-pMHC and LFA1-ICAM1 protein complexes. These segregate to form
a special pattern, known as the immunological synapse (IS), consisting
of a central quasi-circular domain of TCR-pMHC bonds surrounded by a
peripheral domain of LFA1-ICAM1 complexes. Insights gained from
imaging studies had led to the conclusion that the formation of the
central adhesion domain in the IS is driven by active (ATP-driven)
mechanisms. Recent studies, however, suggested that passive
(thermodynamic) mechanisms may also play an important role in this
process. Here, we present a simple physical model, taking into account
the membrane-mediated thermodynamic attraction between the TCR-pMHC
bonds and the effective forces that they experience due to ATP-driven
actin retrograde flow and transport by dynein motor proteins. Monte
Carlo simulations of the model exhibit a good spatio-temporal
agreement with the experimentally observed pattern evolution of the
TCR-pMHC microclusters. The agreement is lost when one of the
aggregation mechanisms is ``muted'', which helps to identify the
respective roles in the process. We conclude that actin retrograde
flow drives the centripetal motion of TCR-pMHC bonds, while the
membrane-mediated interactions facilitate microcluster formation and
growth. In the absence of dynein motors, the system evolves into a
ring-shaped pattern, which highlights the role of dynein motors in the
formation of the final concentric pattern.  The interplay between the
passive and active mechanisms regulates the rate of the accumulation
process, which in the absence of one them proceeds either too quickly
or slowly.

\end{abstract}

\maketitle


\section{Introduction}
\label{sec:intro}

The adaptive immune system heavily relies upon the ability of T cells
to properly interact with antigen presenting cells (APCs). The contact
area between the two cells is established by specific receptor-ligand
bonds that crosslink the plasma membranes of the T cell and the
APC. The key players in this cellular recognition process are T cell
receptor (TCR) and lymphocyte function-associated antigen 1 (LFA1)
that respectively bind to peptide displaying major histocompatibility
complex (pMHC) and intercellular adhesion molecule 1 (ICAM1) embedded
in the APC's plasma membrane~\cite{Alberts,Davis}. During T cell
activation, these two types of bonds are redistributed and form a
unique geometric pattern of concentric supra-molecular activation
centers (SMACs) within approximately 15-30 minutes of the initial
contact~\cite{Grakoui,Johnson}. In this special arrangement, which is
commonly referred to as the immunological synapse (IS), TCR-pMHC bonds
are concentrated into a central SMAC (cSMAC), while the LFA1-ICAM1
adhesion bonds form a surrounding ring termed the peripheral SMAC
(pSMAC)~\cite{Monks,Bromley}. This molecular redistribution is thought
to play an important role in signal regulation~\cite{Mossman}, T cell
proliferation~\cite{Storim}, and focalized secretion of lytic granules
and cytokines~\cite{Griffiths}.

Extensive research effort has been devoted to understanding the
mechanisms governing the formation of the special architecture of the
IS. Mounting evidence from experimental studies point to the actin
cytoskeleton as a vital element in controlling the centripetal motion
of TCR-pMHC and LFA1-ICAM1 bonds towards their final
locations~\cite{Dustin}. In the early stages of IS formation, the T
cell's actin meshwork reorganizes such that actin-free and actin-rich
zone emerge that, later on, constitute the locations of the cSMAC and
pSMAC of the IS, respectively.~\cite{Ritter}. Moreover, actin
polymerization occurring at the periphery of the contact area results
in centripetal actin retrograde flow that is crucial for protein
translocation~\cite{Yi,Babich}. It has been hypothesized that actin
retrograde flow produces viscous forces on the intracellular part of
TCRs, which lead to their centripetal
motion~\cite{Kaizuka,Groves,Wulfing,Hartman}. Furthermore, directed
transport by dynein motor proteins along the cytoskeleton microtubules
has been identified as another ATP-driven process contributing to
protein localization in the IS~\cite{Schnyder,Hashimoto-Tane}.  These
experimental evidences have led to the notion that IS formation is
governed by active cellular processes related to the cytoskeleton
activity.

Interestingly, several theoretical studies have suggested that passive
(non-active) mechanisms may also be involved in the formation process
of the IS~\cite{Chakraborty,Mahadevan,Figge}. These include (i)
receptor-ligand cooperative binding kinetics, and (ii)
membrane-mediated attraction. The former is linked to the fact that
receptor-ligand binding can only occur when the inter-membrane
separation matches the bond length and, thus, new bonds are more
likely to form in regions that are already populated by other
bonds~\cite{Krobath}. The latter is a mechanism of attraction between
distant bonds. It is linked to the reduction in fluctuation entropy
and increase in curvature elasticity that the membranes experience due
to their attachment by the adhesion bonds~\cite{Pincus}. The free
energy cost is minimized when the adhesion bonds are localized in a
single domain which is, therefore, thermodynamically
favorable. Specifically to the IS, TCR-pMHC and LFA1-ICAM1 bonds are
expected to segregate into two distinct domains due to the marked
difference in bond length between them (~$\simeq15\,{\rm nm}$
and~$\simeq42\,{\rm nm}$, respectively~\cite{Lee}), which causes
strong membrane deformation. The interplay between both of these
passive mechanisms may result in patterns that are not only extremely
similar to those observed experimentally, but also form on
biologically relevant timescales~\cite{Chakraborty,Mahadevan,Figge}.

From a physical perspective, domain formation induced by
membrane-mediated attraction can be viewed as a demixing phase
transition~\cite{WeiklReview,WF,Speck2012}.  In a previous study, we
analyzed the membrane-mediated interactions between adhesion bonds and
demonstrated that TCR-pMHC can phase separate from LFA1-ICAM1 and form
domains with similar densities to those in the
IS~\cite{Dharan}. However, conventional phase separation theories are
insufficient to explain the bullseye pattern of the IS, i.e., the
aggregation of TCR-pMHC bonds at the central contact area and the
accumulation of LFA1-ICAM1 bonds at the periphery.  This very
particular structure seems to be directly linked to the activity of
the actin cytoskeleton, especially to directed transport of TCR-pMHC
by dynein motors along microtubules, and the actin retrograde flow
which induces a centripetal force on the TCR-pMHC
bonds~\cite{Burroughs,Weikl2004}. It may also be related to the
depletion of actin from the center of the contact area, which occurs
at the very beginning of the IS formation process. The cytoskeleton is
also expected to have a direct influence on the membrane-mediated
interactions between the TCR-pMHC bonds. This follows from the
attachment of the T cell's membrane to the actin cytoskeleton by
various molecules, such as proteins from the ezrin-mesoin-radixin
(ERM) family~\cite{Shaw}, phosphatidylinositol 4,5- bisphosphate
(PIP2)~\cite{Rodgers,Sun} and coronin 1~\cite{Pieters}. This coupling
between the membrane and the cytoskeleton may modify the shape of the
membrane. Here, we extend our previous analysis of the role played by
membrane-mediated interactions in formation of TCR-pMHC domains, by
including the aforementioned cytoskeleton-related effects, and
studying the interplay between the passive and active mechanisms.

\begin{figure}[t]
\begin{center}
{\centering\includegraphics[width=0.5\textwidth]{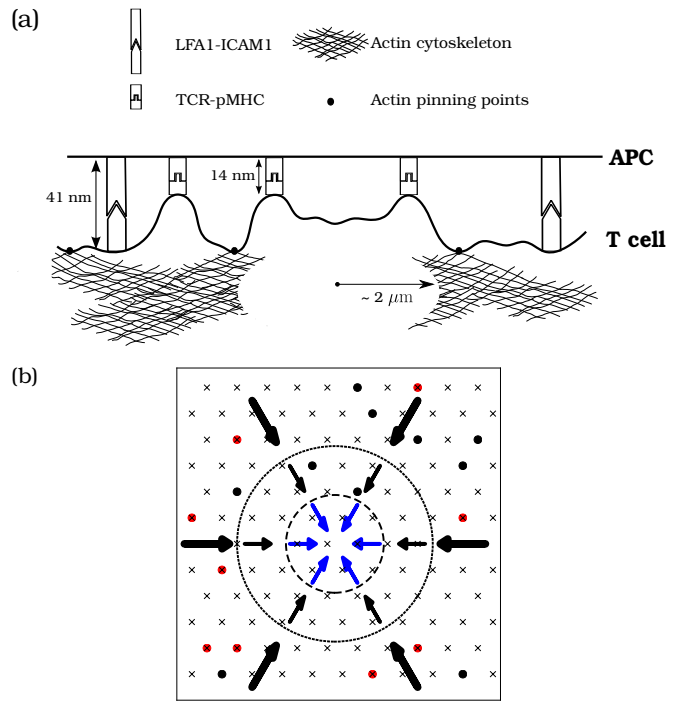}}
\end{center}
\vspace{-0.5cm}
\caption{(a) Schematic of the contact area between the membranes of
  the T cell and the APC. The two membranes are connected by two types
  of adhesion proteins: LFA1-ICAM1 and TCR-pMHC with bond lengths of
  $41\,$ nm and $14\,$nm, respectively. The T cell's membrane is
  attached to the cytoskeleton by a set of actin pinning proteins. A
  central region in the contact area of radius $\sim 2\,{\rm \mu m}$
  is devoid of actin. (b) The lattice model of the contact area
  depicted in (a). The lattice sites are either empty (in which case
  they are marked by the x symbols), or occupied by a single TCR-pMHC
  bond (black circles) or by a single actin pinning point (red
  circles). The latter are immobile and are excluded from the
  actin-depleted central region of the system. The inner dashed circle
  marks the edge of the central actin-depleted region with radius
  $R_C=2\,{\rm \mu m}$, while the outer dotted circle marks the edge
  of the pSMAC, and has a radius $R_P=4\,{\rm \mu m}$. The black and
  blue arrows represent effective centripetal forces arising from the
  actin retrograde flow and directed transport by dynein motors,
  respectively.  For $r < R_P$, the magnitude of the active
  centripetal force is set to $f_0=0.1\,$pN (small arrows), while for
  $r > R_P$ the force is set to a twice larger value of $2f_0 =
  0.2\,$pN, and is indicated by large arrows.}
\label{fig:fig1}
\end{figure}

A schematic picture of the contact area between the APC and the T cell
is depicted in Fig.~\ref{fig:fig1}a, showing the two types of adhesion
bonds that connect them (the longer LFA1-ICAM1 and shorter TCR-pMHC
bonds), the actin cytoskeleton of the T cell including the
actin-depleted circular area of radius $\sim 2\,{\rm \mu m}$, and the
proteins that connect the actin cytoskeleton to the T cell
membrane. In order to follow aggregation of the TCR-pMHC central
domain, we develop a simple lattice model that constitutes a discrete
representation of the system. The lattice model is shown schematically
in Fig.~\ref{fig:fig1}b. It includes three types of sites: (i) empty
(represented by x in Fig.~\ref{fig:fig1}b), or singly occupied by
either (ii) a mobile point representing a TCR-pMHC bond (black circles
in Fig.~\ref{fig:fig1}b), or by (iii) an immobile point representing
an attachment protein between the membrane and the cytoskeleton (red
circles in Fig.~\ref{fig:fig1}b). The latter are absent from the
actin-depleted central region of the contact area (marked by the
dashed circle in Fig.~\ref{fig:fig1}b). The membrane-mediated
potential of mean force (PMF) between the TCR-pMHC bonds is accounted
for via a nearest-neighbor attraction between the mobile points. The
cell cytoskeleton is not modeled explicitly in our coarse-grained
simulations, but is implicitly introduced via an effective potential
that generates the active cytoskeleton forces. These centripetal
forces are represented in Fig.~\ref{fig:fig1}b by arrows, and their
magnitudes are evaluated in section \ref{subsec:flow}. Since we focus
on the aggregation dynamics of TCR-pMHC bonds, we do not study the
very rapid remodeling process of the actin cytoskeleton (which is
completed within less than a minute from the initial contact between
the T cell and the APC~\cite{Ritter}), but consider a system where a
central actin-depleted region has already been formed.

From the simulation results we conclude that the combination of
membrane-mediated attraction, actin retrograde flow, and transport by
dynein motors is essential for proper spatio-temporal evolution of the
TCR-pMHC bonds. We arrive at this conclusion by performing simulations
of systems subjected to only one or two of these driving forces. When
only the active cytoskeleton forces are present, the TCR-pMHC bonds
individually navigate their way through the immobile pinning points
towards the central actin-depleted area. The accumulation process,
however, is completed within an extremely short time of several
seconds, and does not exhibit formation of peripheral TCR
microclusters at early stage. TCR microclusters play a
  vital role in the T cell response prior to the formation of the
  cSMAC that triggers TCR down-regulation and signal
  termination~\cite{Varma,Molnar,DD}. TCR microclusters are observed
in simulations taking into account the membrane-induced attraction but
neglecting the actin retrograde flow and dynein activity; however, in
this case, the aggregation process becomes very slow and proceeds over
many hours. In simulations including the membrane-mediated effect and
actin retrograde flow but missing the dynein motors, the system
evolves into a ring-shaped pattern at the edge of the actin-depleted
area. Similar patterns have indeed been observed in a recent
experimental study where dynein activity was
impaired~\cite{Hashimoto-Tane}. Thus, our results corroborate the
conclusion that actin retrograde flow drives the centripetal motion of
TCR microclusters from the periphery to the center, while dynein motor
proteins govern their translocation at the actin-depleted central zone
and, therefore, play an essential role in the formation of the final
circular bullseye pattern of the IS. When all three mechanisms are
present, the central accumulation of the TCR-pMHC bonds occurs within
about half an hour, which is indeed the correct biological
timescale. The simulations do not only reproduce the experimentally
observed values for the timescale of IS formation, but are also able
to capture two critical features of this process, namely, (i) the
peripheral coarsening of TCR-pMHC into microclusters, and (ii) their
subsequent transport towards the center of the contact region. From
the simulations we conclude that membrane elasticity and active
cytoskeleton forces act as complementary mechanisms that balance each
other in a manner that assures an adequate T cell response.

\section{Model and simulations}
\label{sec:model}

\subsection{Lattice-gas model}
\label{subsec:LG}

The accumulation of TCR-MHC bonds at the center of the contact area
between a T cell and an APC is studied using a lattice-gas model. In
this coarse-grained physical framework, the cell membranes are
implicitly accounted for by nearest neighbor interactions that
represent the PMF originating from the membrane deformation energy. We
consider a triangular lattice with lattice spacing $\xi=100\,$nm,
which is the range of the membrane-mediated interactions in the system
under consideration here (see supplementary information SI1). This
sets the spatial resolution of out model, and allows us to ignore
direct (e.g., van der Waals and diploar~\cite{Israelachvili}) and
lipid-mediated~\cite{Gomez} interactions between the various proteins
in the system since, typically, the range of these interactinos does
not extend beyond $\lesssim10\,$nm. The linear size of the system is
roughly $L=10\,{\rm \mu m}$ (we simulate a lattice of $99\times114$
sites with an aspect ratio close to unity), which is representative of
the dimensions of the contact area. The model includes two types of
lattice points representing the TCR-pMHC bonds (type A) and the
membrane-cytoskeleton pinning proteins (type B). The former are
mobile, while the latter are located at fixed lattice sites. Both A-A
and A-B interactions are purely repulsive at short separations (see
SI1), and this feature is accounted for by prohibiting multiple
occupancy of a lattice site.  At a distance $\xi$, the A-A
interactions are attractive (SI1), and this is represented in the
model via a nearest neighbor interaction energy of strength
$\epsilon=-4.5~k_{\rm B}T$. The value of $\epsilon$ is derived from a
statistical-mechanical calculation for the membrane-mediated PMF
between two TCR-pMHC bonds (SI1). The model does not include explicit
representation of the LFA1-ICAM1 bonds. The effect of these bonds is
incorporated in the statistical-mechanical calculations of the PMFs
through the parameter $h_0=27\,$nm (SI1), which is the bond length
mismatch between LFA1-ICAM1 and TCR-pMHC bonds 
  \cite{footnote1}.

The densities of the TCR-pMHC bonds (type A lattice points) and the
membrane-cytoskeleton pinning proteins (type B) greatly vary between
different experimental works (if reported at all). We therefore set
their values in the simulations based on the following considerations:
Type A points aggregate to form the cSMAC, which is a circular domain
of radius $\sim 2\,{\rm \mu m}$ that almost fills (at the end of the
process) the actin-depleted central region. The latter includes about
1200 lattice sites, and the number of type A points is set to a
slightly smaller value $N_A=1000$.  We note that since the lattice
spacing $\xi=100\,$nm, the density of the TCR-pMHC bonds in a cluster
is $\simeq80~{\rm bonds/ \mu m^2}$, which is indeed above the
threshold density required for full T cell
activation~\cite{Grakoui,Bromley}. A reasonable estimate to the
spacing, $l$, between membrane-cytoskeleton pinning proteins is a
distance of a few hundred nm~\cite{Kusumi}. In what follows, we show
results for systems where the average spacing between type B points is
set to $l=300\,$nm. To avoid further complication of our model, we
neglect the binding/unbinding kinetics of the membrane-cytoskeleton
linkers and assume that the number of type B points is
fixed. Nevertheless, we note here that results of simulations with
$l=500\,$nm (i.e., with fewer type B points) exhibit negligible
differences.  The type B points can be found everywhere on the
lattice, except for the central actin-depleted region. They are
randomly distributed to account for local variations in $l$, but we do
not allow two type B points to occupy nearest-neighbor lattice sites
in order to ensure a fairly uniform distribution and avoid type B
clusters.

\subsection{Monte Carlo simulations}
\label{subsec:MC}

We perform Monte Carlo (MC) simulations to study the evolution of the
lattice model. While MC simulations are designed for statistical
ensemble sampling at equilibrium, they can be also used to effectively
generate Brownian dynamics in lattice models. The simulations consist
of move attempts of a randomly chosen type A point to a nearest
lattice site, which is accepted according to the standard Metropolis
criterion~\cite{Metropolis}. During a single MC time unit, $\tau_{\rm
  MC}$, each type A point experiences (on average) one move
attempt. Mapping the MC time unit to real time, $t$, can be achieved
by considering the two-dimensional diffusion relation $\langle
r^2\rangle=4Dt$, where $\langle r^2\rangle$ denotes the mean square
displacement and $D$ is the diffusion coefficient. For the MC
simulations, each point moves a distances of the lattice spacing
$\xi=100\,$nm and, therefore, $\tau_{\rm MC}=\xi^2/4D$. This can be
compared to typical values for the diffusion coefficient of T cell
membrane proteins $D\simeq0.1~{\rm \mu m^2/sec}$~\cite{Favier}, which
yields $\tau_{\rm MC}\simeq 25$~msec.

\subsection{Active cytoskeleton forces}
\label{subsec:flow}

Similarly to the approach taken in ref.~\cite{Weikl2004}, active
cytoskeleton processes are not modeled explicitly in our simulations
but are instead represented by the effective forces that they induce
on the TCR-pMHC bonds. Two active processes are considered, namely (i)
actin retrograde flow, and (ii) dynein minus-end directed transport
toward the microtubule organizing center (MTOC). These may be viewed
as complementary mechanisms since they produce the same net effect of
transport toward the center, while operating at different regions of
the system. Explicitly, dynein-driven transport governs the dynamics
at the central actin-depleted area, while actin-retrograde flow is
believed to predominate at the periphery of the system.

Experimental studies reveal that the velocity of the actin retrograde
flow towards the center of the contact area achieves a maximal value
$v_{\rm max}\simeq0.1~{\rm \mu m/sec}$ at the periphery of the contact
area. As the flow proceeds toward the center of the contact area, it
decreases to approximately $0.5v_{\rm max}$, and finally vanishes at
the edge of the central actin-depleted region~\cite{Yi,Kumari}. The
flow generates a centripetal force on the TCR-pMHC bonds with
magnitude proportional to the flow velocity. Inside the actin-depleted
region, the centripetal motion of the TCR-pMHC bonds continues, but is
now driven by the activity of dynein motors. This has been concluded
by experiments showing that in the absence of dynein motor activity,
the TCR-pMHC bonds do not penetrate into the actin-depleted region,
but instead accumulate around it. Modeling the dynein-induced forces
is a highly challenging task since the motor activity depends on many
factors, including the concentration of available ATP, the density of
dynein motors, and the load opposing the motors which may depend on
the size of the transported microcluster. To bypass this issue, we
clump all of these factors together into an effective dynein-driven
force that acts on individual TCR-pMHC bonds inside the actin-depleted
zone, and pulls them towards the center of the system (where the MTOC
is located). Experiments suggest that the centripetal velocity of
TCR-pMHC bonds transported by both actin retrograde flow and dynein
motors is roughly twice larger than the velocity in the absence of
motor activity, which suggests that the effective centripetal force
induced by both mechanisms is of the same order of magnitude~
\cite{Hashimoto-Tane}.  Taking these various considerations into
account, the combined active cytoskeleton forces are introduced to the
model via the following effective potential that depends on the
distance $r$ from the center of the system
\begin{equation}
\label{eq:active}
\Phi_{\rm active}(r)=\left\{\begin{array}{lc}
f_0\,r & r\le R_{\rm P} \\
2f_0\left(r-R_{\rm P}\right)+f_0R_{\rm P}\ & r>R_{\rm P},
\end{array}\right.
\end{equation}
where $R_P$ is the outer radius of the pSMAC, which is the region
where the flow of actin becomes weaker~\cite{Kumari}. This potential
generates a centripetal force of magnitude $f_0$ for $r\leq R_{\rm P}$
and $2f_0$ for $r\geq R_{\rm P}$. The former region includes the
actin-depleted region ($0<r<R_{\rm C}$) and the pSMAC ($R_{\rm
  C}<r<R_{\rm P}$), where a force of magnitude $f_0$ is induced by
dynein motor activity and actin retrograde flow, respectively. The
latter region corresponds to the periphery of the cell, where actin
retrograde flow is stronger and produces an effective force of
magnitude $2f_0$. A schematic depicting the forces associated with the
potential $\Phi_{\rm active}$ at different regions of the contact area
is shown in Fig.~\ref{fig:fig1}b. Based on confocal images, we set
$R_{\rm C}=2\,{\rm \mu m}$ and $R_{\rm P}=4\,{\rm \mu
  m}$~\cite{Kaizuka}. In order to determine the value of $f_0$, we use
the observation that the actin retrograde flow causes a small
peripheral microcluster (a few hundred nanometers in size) to move
toward the cell's center at a velocity of $v_{c}\simeq20\,{\rm
  nm/sec}$~\cite{DeMond}. We, thus, performed short MC simulations for
a single microcluster composed of 10 TCR-pMHC bonds, which is located
at the outer region of the system. We measured the velocity of the
centripetal movement of the microcluster as a function of $f_0$, and
found that it attains the value of $v_c$ for $f_0=0.1\,{\rm
  pN}$. Comparable forces have been measured in experiments of
optically trapped microbeads coupled to actin retrograde flow of
similar velocity~\cite{Mejean}.

\section{Results}
\label{sec:results}

\begin{figure}[t]
\begin{center}
{\centering\includegraphics[width=0.5\textwidth]{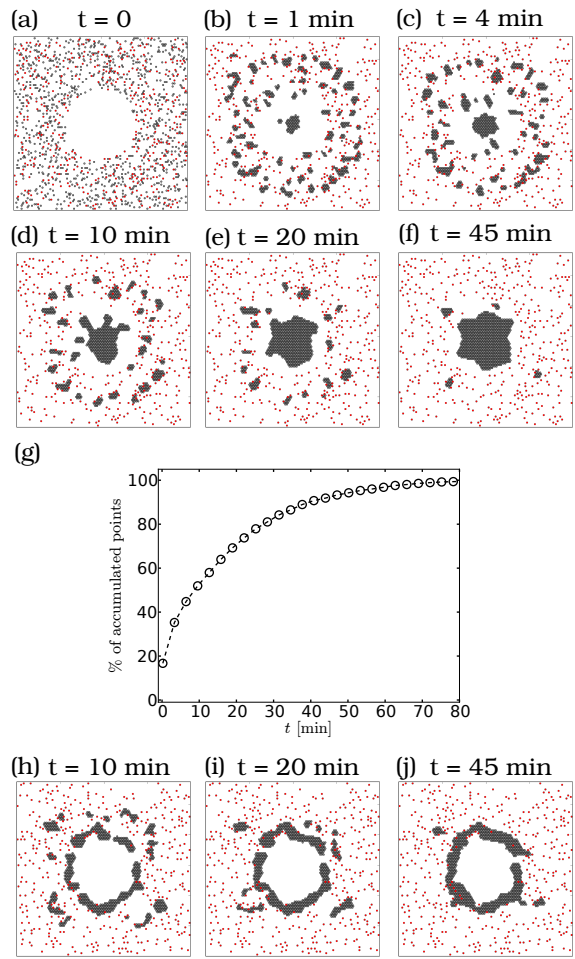}}
\end{center}
\vspace{-0.5cm}
\caption{Simulation snapshots depicting the aggregation process of the
  TCR-pMHC bonds (type A, black dots) at different times: (a) The
  random initial distribution, (b-d) early stage formation,
  coarsening, and centripetal drift of microclusters, and (e,f) late
  stage accumulation of microclusters and cSMAC formation. The red
  dots represent the cytoskeleton pinning proteins (type B). (g) The
  percentage of type A bonds located at the central actin-depleted
  area as a function of time. (h-j) The evolution of the system in the
  absence of dynein forces at the center.}
\label{fig:fig2}
\end{figure}

The spatio-temporal evolution of the system has been analyzed from 10
independent MC runs starting with different random distribution of
both type A and B points. Typical snapshots from different stages of
the process are shown in Fig.~\ref{fig:fig2}, with the type A and B
points presented by black and red dots, respectively. At $t=0$, the
type A points are randomly distributed outside of the inner
actin-depleted regime (Fig.~\ref{fig:fig2}a).  Within less than one
minute, the type A points coalesce and form small peripheral
microclusters consisting of $\lesssim20$ points
(Fig.~\ref{fig:fig2}b), which begin to move centripetally. Several
type A points have already reached the center system at this
stage. The peripheral microclusters are believed to play a vital role
in initiating and sustaining TCR signaling~\cite{Varma,Campi}.  As
time proceeds, the clusters further coarsen, which decreases their
mobility and slows down their centripetal motion
(Figs.~\ref{fig:fig2}c-e). The increase in the size of the clusters
also causes them to be more affected by the presence of the type B
points which act as repelling obstacles. This further restrains the
centripetal movement of the microclusters, which have to ``navigate''
their way through the ``curvature corrals'' that the type B points
form. We also observe in Figs.~\ref{fig:fig2}c-e the gradual increase
in the size of the central domain, which is the destination where the
microclusters accumulate. After 45 minutes (Fig.~\ref{fig:fig2}f),
almost all type A points reside in the central domain. The dynamics of
the MC simulations, as depicted in Figs.~\ref{fig:fig2}a-e, closely
resembles epifluorescence and total internal reflection fluorescence
(TIRF) microscopy images of the IS formation process. Specifically,
the microscopy images show the generation of similar microclusters,
their drift to the center of the contact area, and their accumulation
at the center of the contact area~\cite{Hartman,Yokosuka}. The
simulations exhibit a very good agreement with the experimental
observations not only with regard to spatial evolution of the system,
but also with respect to the time scales of the different stages of
the process.  Fig.~\ref{fig:fig2}g depicts the percentage of type A
points located at in the central domain at the actin-depleted
area. About 90\% of the lattice points have been accumulated at the
center after roughly 40 minutes, which agrees very well with the times
reported in the literature for cSMAC formation.

Figs.~\ref{fig:fig2}h-j show snapshots from simulations in which
dynein activity is turned off. This is done by modifying the effective
potential (\ref{eq:active}) such that $\Phi_{\rm active}\left(r<R_{\rm
  C}\right)=f_0\,R_{\rm C}={\rm Constant}$, and thus, the associated
force vanishes at the actin-depleted area where the dynein motors
operate. For this model system, we observe formation of peripheral
microclusters that move centripetally, but do not enter the
actin-depleted area. Instead of forming a central quasi-circular
domain, the type A points now accumulate in a ring-shaped domain at
the edge of the actin-depleted region. Interestingly, very similar
ring-like structures have been observed in experiments where dynein
activity was inhibited by dynein heavy chain
ablation~\cite{Hashimoto-Tane}. From the agreement with the
experimental results we conclude that cSMAC formation requires that
centripetal forces act on the TCR-pMHC bonds in the actin-depleted
area, and that the origin of these forces is the action of the dynein
motors.
 
\begin{figure}[t]
\begin{center}
{\centering\includegraphics[width=0.5\textwidth]{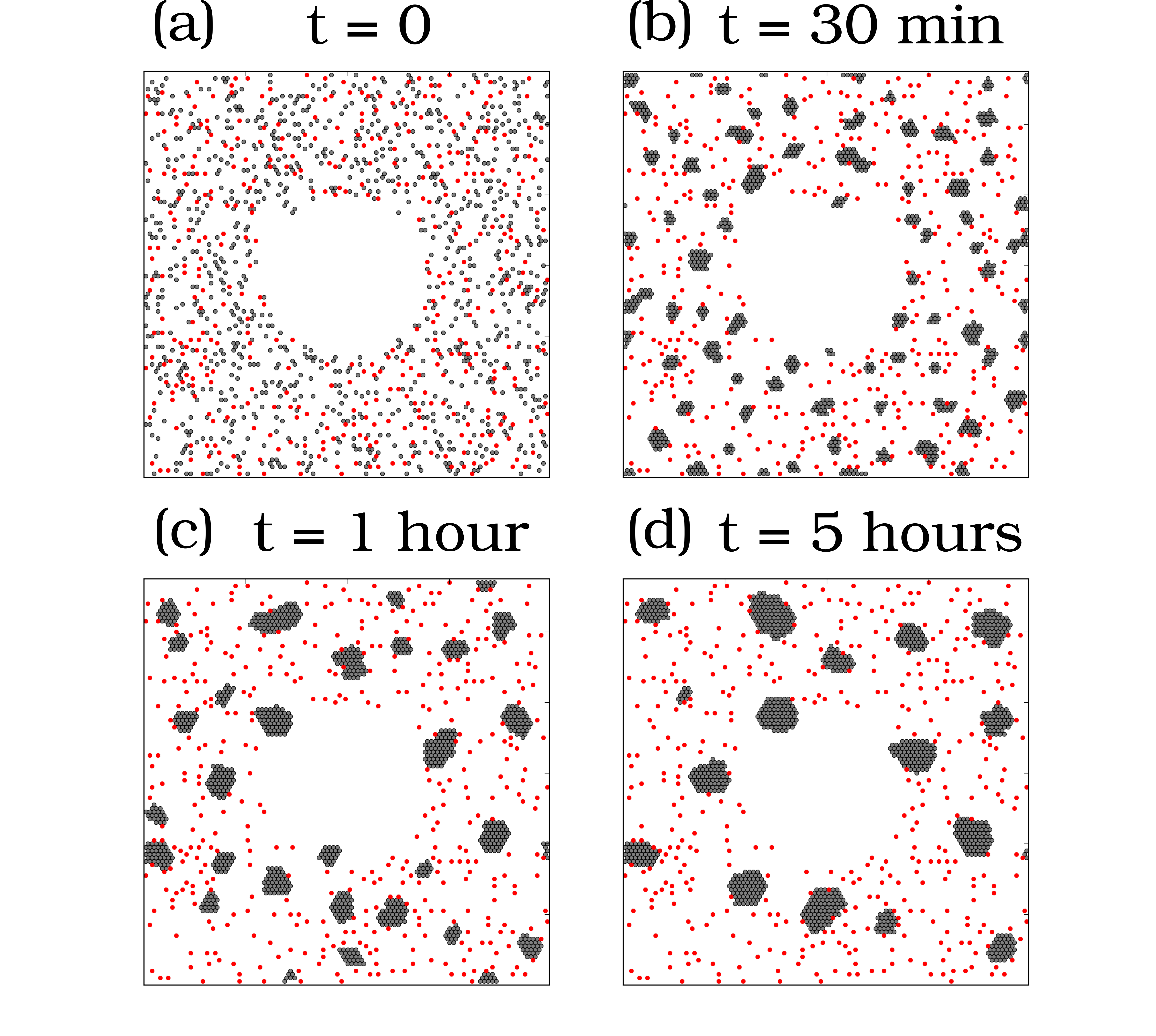}}
\end{center}
\vspace{-0.5cm}
\caption{Simulation snapshots depicting the aggregation process in the
  absence of active cytoskeleton forces. The type A points form
  microclusters that, on time scales of hours, barely grow and do not
  exhibit a drift toward the central area. Color coding as in
  {\protect Fig.~\ref{fig:fig2}}}
\label{fig:fig3}
\end{figure}

\begin{figure}[t]
\begin{center}
{\centering\includegraphics[width=0.5\textwidth]{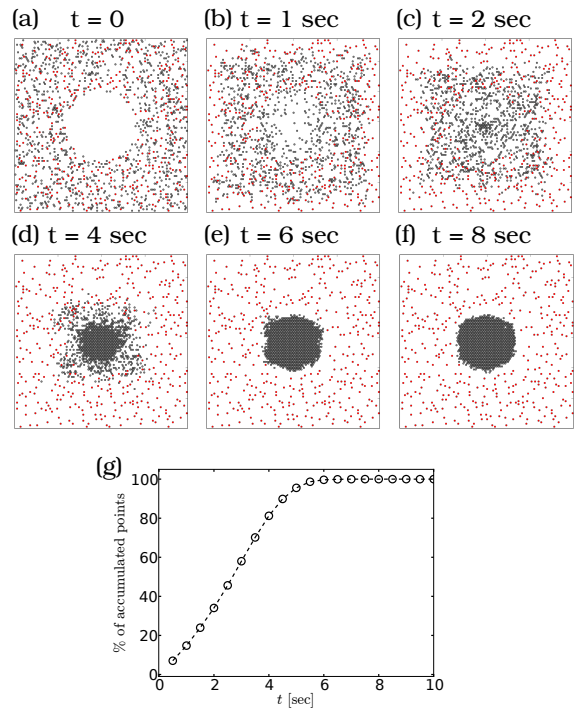}}
\end{center}
\vspace{-0.5cm}
\caption{Simulation snapshots depicting the aggregation process in the
  absence of membrane-mediated attraction. The type A points do not
  form microclusters, and accumulate at the central area within a few
  seconds. Color coding as in {\protect Fig.~\ref{fig:fig2}}. (g) The
  percentage of type A bonds located at the central actin-depleted
  area as a function of time.}
\label{fig:fig4}
\end{figure}

The results displayed in Fig.~\ref{fig:fig2} appears to be in
agreement with the widely-held view that the active cytoskeleton
forces due to actin retrograde flow and the dynein motors determine
the final destination of the TCR-pMHC bonds, which is the center of
the cell when dynein activity is enabled and the inner edge of the
actin-rich zone when it is disabled. The results also demonstrate that
while a central domain located inside the actin-depleted area
constitutes the equilibrium state of the system from a pure
thermodynamic perspective, membrane-mediated interactions do not
contribute significantly to the centripetal accumulation of the TCR
microclusters over the biologically relevant time scales. This is
demonstrated in Fig.~\ref{fig:fig3}a-d showing snapshots from
simulations where $\Phi_{\rm active}$ is completely turned off,
leaving the system to evolve only under the influence of the passive
membrane-mediated interactions. Neither centripetal motion nor central
accumulation of TCR-pMHC bonds are observed in this set of
simulations. However, the data seems to indicate that the
membrane-mediated interactions play an important role in facilitating
the formation and coarsening of peripheral TCR microclusters, which
are known to be essential for adequate T-cell immune response. The
importance of the passive interactions in inducing the formation of
the TCR microclusters is illustrated in Fig.~\ref{fig:fig4}, showing
snapshots from yet another set of simulations. Here, the
nearest-neighbor membrane-mediated attraction between the type A
points is muted by setting $\epsilon=0$, while the effective
centripetal potential $\Phi_{\rm active}$ (\ref{eq:active}) is
retained. The simulations reveal that, in this case, the type A points
do not form microclusters but instead move individually and quickly
accumulate at the center of the system.  Since the mobility of a
single point is higher than that of a cluster, it is expected that the
aggregation process is completed in a shorter time than required in
the presence of membrane-mediated attractions.  Our computational
results show that, indeed, 100\% of the points arrive at the central
area in less than 6 seconds (see Fig.~\ref{fig:fig4}g).  We note here
that the rate of the accumulation process depicted in
Fig.~\ref{fig:fig4} may be exaggerated, since it implies that
individual (non-clustered) TCR-pMHC move centrally at a velocity which
is about 5 times larger than the velocity of the actin retrograde
flow. This problematic dynamic feature is an artifact of the lattice
MC dynamics that we employ. Since the focus of interest is the
evolution of the system over durations of minutes and hours, the actin
retrograde flow force has been calibrated to produce correctly the
centripetal velocity of small clusters on these temporal scales - see
detailed discussion in section~\ref{subsec:flow}. The consequence of
this choice is that the velocity of the individual TCR-pMHC bonds
turns out to be too high. A reasonable estimation for the centripetal
velocity of individual bonds at the periphery of the system is
$20<v<100~{\rm nm/sec}$, i.e., higher than the velocity of a single
small microcluster but lower than the actin retrograde flow
velocity. The accumulation time of bonds moving at such speeds is
0.5-2 minutes, which is still more than an order of magnitude smaller
than the IS formation time. Comparing the results of
Fig.~\ref{fig:fig4} from simulations that miss the membrane-mediated
attraction with Fig.~\ref{fig:fig2} highlights two important aspects
of membrane elasticity: First, it facilitates early microcluster
formation that play a vital role in T cell activation. Second, they
lead to coarsening dynamics that result in a decrease in the mobility
of the clusters and, thereby, regulates the duration of the
aggregation process.

\section{Discussion and summary}
\label{discuss}

The formation of the immunological synapse (IS) is a complex
biological process involving multiple molecular components, including
several adhesion proteins, motor proteins, the actin cytoskeleton, and
the membranes of both the T cell and the antigen presenting
cell. Adhesion between the two cell membranes is established by two
types of receptor-ligand bonds, namely, TCR-pMHC and LFA1-ICAM1. At
the onset of the process, the T cell's cytoskeleton remodels and an
actin-depleted region is formed at the center of the contact area
between the cells.  Within several tens of minutes, the adhesion bonds
segregate such that the TCR-pMHC bonds aggregate into a quasi-circular
domain located at the actin-depleted region. The macroscopic time
evolution of such a complex system can be only addressed through
coarse-grained models employing simplified molecular representation
and focusing on the most dominant biophysical mechanisms. Here, we
present a minimal computational lattice model aiming to study the
dynamics of the TCR-pMHC bonds (represented as lattice points). The
model takes into account several forces that emerge in the literature
as key factors in TCR-pMHC localization.  One is a passive
thermodynamic force associated with the membrane curvature energy,
which induces attraction between the TCR-pMHC bonds, and provides
repulsion between the TCR-pMHC bonds and the proteins connecting the
actin cytoskeleton to the membrane. The others are centripetal forces
exerted on the TCR-pMHC bonds due to the actin retrograde flow and
directed transport by dynein motors along microtubules. These are
active (non-equilibrium) forces because the processes that they
originate from are driven by consumption of ATP chemical energy.

Our coarse-grained model and MC simulations produce correct evolution
dynamics of the IS formation process at the experimentally observed
time scales. They thus provide a clear and intuitive picture for the
roles played by the main driving forces, and into the intricate
interplay between them. Membrane-mediated attraction facilitates the
formation of TCR-pMHC microclusters. These microclusters, which
initiate biological cues necessary for T cell activation, are rapidly
formed and continue to grow by coarsening dynamics over a period of
approximately 10 minutes. The TCR-pMHC microclusters are corralled by
the membrane-cytoskeleton binding proteins of the T cell, with which
they have repulsive membrane-mediated interactions. As the
microclusters grow in size, the corralling effect becomes more
significant and their diffusivity decreases. This inhibits their
accumulation in a central quasi-circular domain, despite the fact that
this configuration represents the equilibrium state of the
system. What speeds up the dynamics of the microclusters and directs
their movement toward the center of the contact area is a centripetal
active force induced by actin retrograde flow at the periphery of the
synapse. This observation corroborates the largely consensual view
about the importance of actin remodeling and retrograde flow for the
centripetal translocation of TCR-pMHC bonds at the IS. It is, however,
important to emphasize that in simulations without membrane-mediated
attraction, a central domain forms very rapidly without exhibiting
intermediate microclusters. This observation highlights the, rather
overlooked, important role played by the membrane-mediated
interactions in regulating the rate of the IS formation process. This
novel conclusion drawn from our model and simulations has not been
addressed experimentally thus far. A possible setup by which it can be
tested is a model system consisting of all the molecular ingredients
of the system, except for the LFA1-ICAM1 bonds. This would shut down
the membrane-mediated interactions whose origin is is the mismatch in
bond length between the TCR-pMHC and the LFA1-ICAM1 bonds.

At the central actin-depleted region, a centripetal force (of
magnitude similar to the actin retrograde flow induced force) is
effectively generated by dynein motors that walk toward the minus-end
of the microtubules. An interesting observation is the formation of a
ring-shaped domain at the edge of the actin-depleted region in
simulations where this force is turned off. This result is in line
with a recent study, in which similar structures were observed when
dynein activity was genetically impaired, but actin retrograde flow
was maintained. The inability of the system to produce a central
circular domain in the absence of dynein activity highlights the role
played by the motors at the final stage of the process, where they
enable the TCR microclusters to enter the actin-depleted
region. Furthermore, the agreement between our simulation results and
the experimental data supports the notion presented here of
considering two different concentric regions for the active forces,
namely a central area dominated by the dynein motors, and a more
distant one where the actin retrograde flow is the origin of the
centripetal movement of the microclusters.

To conclude, we explored the role played by passive and active forces
in the process of IS, and demonstrated how the interplay between them
regulates the spatio-temporal pattern formation. Only in simulations
where all the driving forces were present, the signature features of
the IS formation process were observed. These include microclusters
formation and coarsening dynamics, their transport to the central
actin-depleted area, and their aggregation into a single
quasi-circular domain (the cSMAC). Moreover, the simulated system
evolves at the experimentally observed rates: Microclusters are formed
within roughly a minute from the beginning of the process, and the
final bullseye structure is formed within 15-30 minutes.  Our results
point to an interesting role of membrane elasticity in regulating the
IS formation process in a manner that provides T cells with an
appropriate time window to allow biological signaling via the
peripheral TCR microclusters, which is vital for a proper immune
response.

This work was supported by the Israel Science Foundation (ISF) through
grant 1087/13.



\pagebreak
\
\pagebreak

\renewcommand \thesection{\Roman{section}}
\renewcommand{\thefigure}{S\arabic{figure}}
\setcounter{figure}{0}

{\bf \Large Supplementary Information 1}

\vspace{0.5cm}

{\bf Membrane-mediated interactions}

\vspace{0.5cm}

\begin{figure}[tb]
\begin{center}
  {\centering\includegraphics[width=0.5\textwidth]{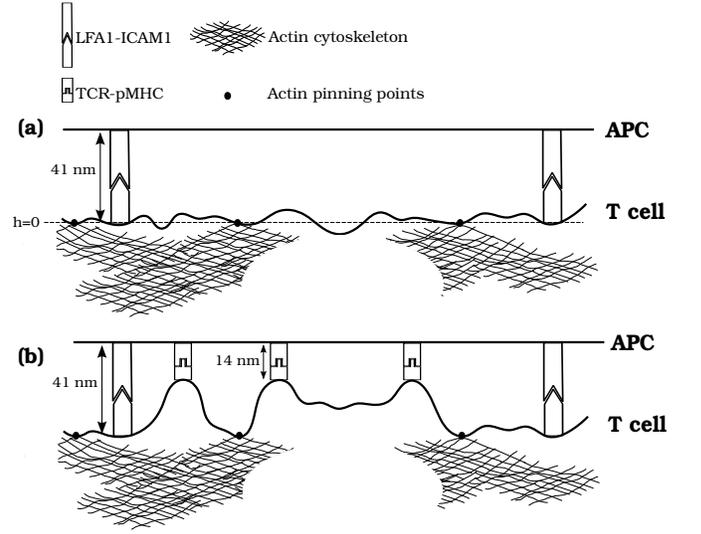}}
\end{center}
\vspace{-0.5cm}
\caption{Schematics of the contact area between the membranes of the T
  cell and the APC. The two membranes are connected by two types of
  adhesion proteins: LFA1-ICAM1 and TCR-pMHC with bond lengths of 41
  nm and 14 nm, respectively. The T cell's membrane is attached to the
  cytoskeleton by a set of actin pinning proteins. (a) shows all of
  the system components except from the TCR-pMHC bonds. The latter,
  which are present in (b), pull the T cell membrane away from its
  reference state.}
\label{fig:figSI1}
\end{figure}

\begin{figure}[tb]
\begin{center}
  {\centering\includegraphics[width=0.5\textwidth]{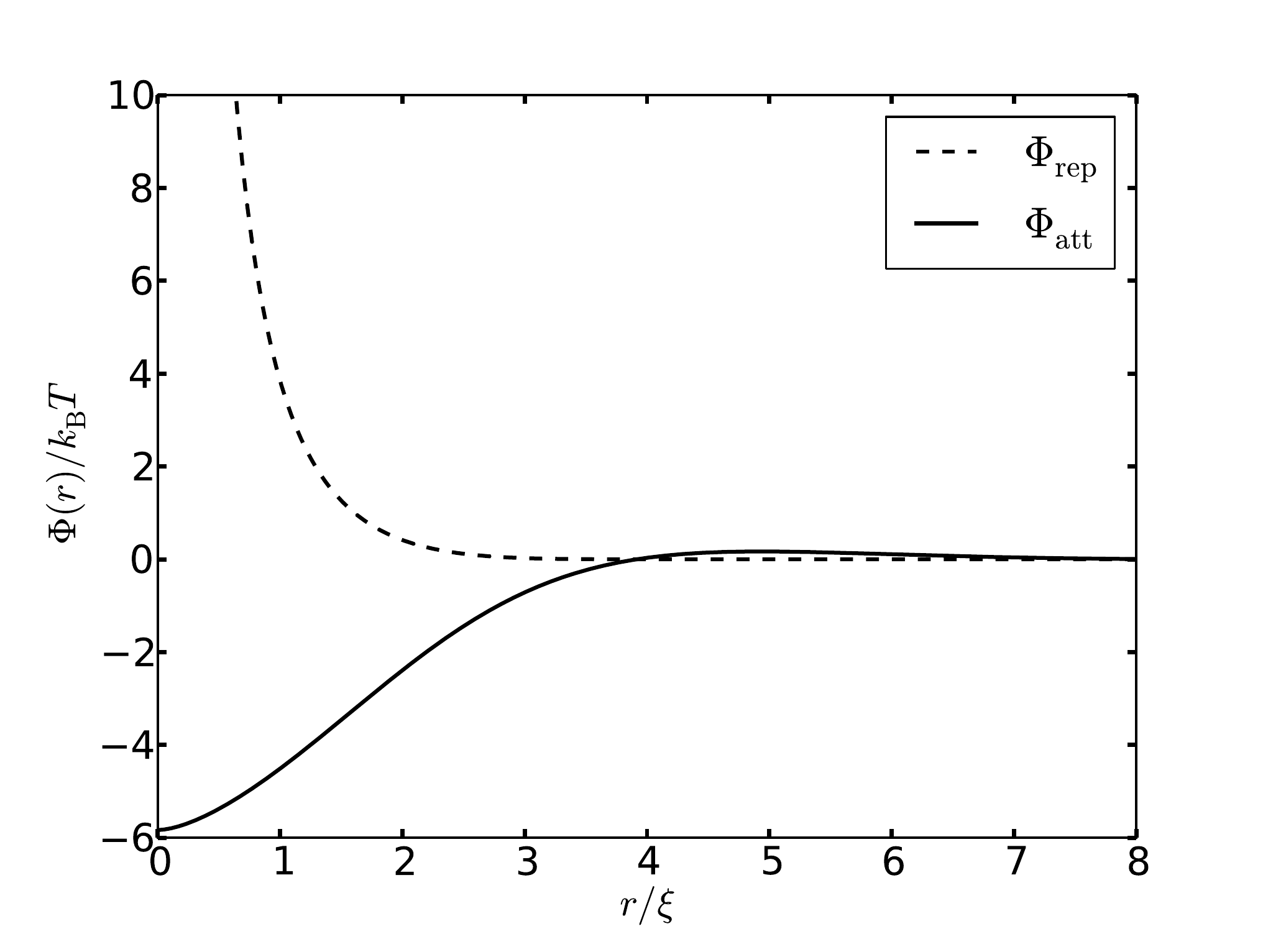}}
\end{center}
\vspace{-0.5cm}
\caption{Curvature-induced interactions: the solid line depicts the
  curavture-induced attraction between two TCR-pMHC bonds, while the
  dashed line stands for the curvature-induced repulsion between a
  TCR-pMHC bond an a membrane-cytoskeleton pinning point.}
\label{fig:figSI2}
\end{figure}

The interaction between a T cell and an APC is mediated by
several molecular components, especially the cells plasma membranes
and the LFA1-ICAM1 and TCR-pMHC adhesion bonds connecting them. In our
simulations, we focus on the dynamics of the TCR-pMHC bonds and for
this purpose we need to calculate the membrane-mediated interactions
that they experience due to the deformation of the membrane of the T
cell. To this end, the state of a system that includes all the
relevant components, {\em except for the TCR-pMHC bonds}\/, is taken
as a reference. In this reference state, which is shown schematically
in Fig.~\ref{fig:figSI1}a, the distance between the cells is set by
the length of the LFA1-ICAM1 bonds ($\sim 41\,$ nm). The deformation
energy of the T cell membrane can be described by the Helfrich
effective Hamiltonian~\cite{Helfrichsi}
\begin{equation}
\label{eq:HH}
{\cal H}=\int\frac{1}{2}\left[\kappa \left(\nabla^2h \right)^2 +
  \gamma h^2\right]d^2{\bf r},
\end{equation}
where $h({\bf r})$ is the membrane's height profile relative to the
reference plain state at $h=0$. The first term in eq.~(\ref{eq:HH}) is
the curvature energy of the membrane, characterized by a bending
modulus $\kappa$. The second term is a harmonic confining potential of
strength $\gamma$. This term accounts for various influences that
surpress the membrane thermal undulation around $h=0$, e.g., the
interaction of the membrane with the actin cytoskeleton (residing
underneath the T cell membrane in Fig.~\ref{fig:figSI1}a) and the
glycocalyx (not shown explicitly). Fig.~\ref{fig:figSI1}b depicts the
same system with the TCR-pMHC included. These pull the T cell membrane
toward the APC and locally set the inter-cell separation to 14 nm,
thereby deforming the T cell membrane a distance $h_0=41-14=27$ nm
from the minimum of the harmonic confining potential. The resulting
deformation energy is minimized when the TCR-pMHC bonds aggregate,
which is the origin of the attractive membrane-mediated interactions
between them. In principle, these interactions are described by a
many-body potential of mean force (PMF) that depends on the
instataneous coordinates of all TCR-pMHC bonds. In practice, for
dilute systems the PMF is well-approximated by the sum of pairwise
interactions that depend only on the distance $r$ between the adhesion
bonds~\cite{Speck2010si}. The PMF between two TCR-pMHC bonds, $\Phi_{\rm
  att}(r)$, can be derived from the partition function
\begin{equation}  
\label{eq:Za}
Z_A=\int{\cal D}\left[h({\bf r})\right] e^{-\beta{\cal
    H}}\delta\left(h(0)-h_0\right)\delta\left(h(r)-h_0\right),
\end{equation}
which involves statistical averaging over all membrane configurations
whose height at the locations of the bonds $(r_1=0,r_2=r)$ is fixed at
$h_0=27\,$nm. The height constraints are represented in
eq.~(\ref{eq:Za}) by the two $\delta$-functions. The partition
function can be calculated using standard statistical-mechanics
techniques for handling multidimensional Gaussian integrals. The PMF
is related to $Z_A$ by $\Phi_{\rm att}(r)=-k_{\rm B}T\ln{Z_A}$, where
$k_{\rm B}$ is the Boltzman constant and $T$ is the temperature. It is
given by (see eqs.(5) and (7) in ref.~\cite{Dharansi})
\begin{equation}
\label{eq:F2}
\frac{\Phi_{\rm att}(r)}{k_{\rm B}T}=\left(\frac{h_0}{\Delta}\right)^2
\frac{\frac{4}{\pi}{\rm
    kei}\left(\frac{r}{\xi}\right)}{1-\frac{4}{\pi}{\rm
    kei}\left(\frac{r}{\xi}\right)},
\end{equation}
where ${\rm kei(x)}$ is the Kelvin function~\cite{Abramowitzsi},
$\xi=\left(\kappa/\gamma\right)^{1/4}$ and $\Delta^2=k_{\rm
  B}T/8\sqrt{\kappa\gamma}$. Both $\xi$ and $\Delta $ have units of
length. For T cells, their values are roughly given by
$\xi\simeq100\,$nm and $\Delta\simeq8\,$nm~\cite{Dharansi}. The PMF,
$\Phi_{\rm att}/k_{\rm B}T$ (expressed in units of the thermal
energy), is depicted by the solid line in Fig.~\ref{fig:figSI2} as a
function of the normalized pair distance $r/\xi$, for the
aformentioned values of the systems parameters $\xi$, $\Delta$ and
$h_0$. The attractive pair-potential has a strength of a few $k_{\rm
  B}T$ for $r\simeq \xi$, and is screened at somewhat larger
distances.

At separations smaller than $\xi$, one has to take into account direct
excluded volume (hard core) interactions between the TCR-pMHC
bonds. Since these are missing in the calculation of the partition
function, a purely repulsive potential diverging for $r\to 0$ must be
added to $\Phi_{\rm att}$. The full pair potential between TCR-pMHC
bonds is reminiscent of a Lennard-Jones potential, i.e., repulsive at
very short distances and attractive at an intermediate finite
range. Domain formation under the influence of this type of potentials
can be conviniently studied within the framework of the classical
discrete lattice-gas (Ising) model where each lattice site can be
occupied by at most one lattice point (in order to account for the
short range repulsion), and with nearest-neighbor attraction of
strength $\epsilon$. Setting the lattice spacing to $\xi$, we take
$\epsilon=\Phi_{\rm att}(r=\xi)\simeq-4.5~k_{\rm B}T$. We do not
consider next-nerarest-neighbor interactions, despite of the fact that
the $\Phi_{\rm att}$ does not fully decay at $r=\xi$. The reason for
this decision is the many-body nature of the membrane-mediated PMF,
which becomes important at the onset of the formation of adhesion
clusters. In high density domains, each adhesion bond interacts with
the proximal bonds in the first surrounding shell, whose very presence
screens the interactions with the slightly more distant bonds in the
next shells~\cite{WFsi}.

Another important note about the many-body membrane-mediated PMF
between TCR-pMHC bonds is that, in principal, it should also be
dependent on the density of the LFA1-ICAM1 bonds since their presence
is the source of effective attraction between the TCR-pMHC
bonds. Here, the LFA1-ICAM1 bonds are not modeled explicitly but are
represented by the uniform harmonic potential in
eq.~(\ref{eq:HH}). The utility of this approach was demonstrated in a
previous study~\cite{Dharansi}, employing a lattice model with a
microscopic spacing of 5 nm and an explicit representation of the
membrane. In that work we calculated the phase diagram of the system,
which exhibits a two-phase region including semi-dilute domains with
densities of about 100 bonds per ${\rm \mu m}^2$. These densities are
comparable to the estimated density of TCR-pMHC bonds in the IS, and
correspond to a typical distance of $\xi=100$ nm between them. Thus,
the uniform harmonic confining potential correctly captures the
influence of the LFA1-ICAM1 bonds in inducing the aggregation of the
TCR-pMHC bonds in the system.  Based on the success of
eq.~(\ref{eq:HH}) to capture the thermodynamics of the system within a
model of finer resolution, we here present a coarser model with
lattice spacing $\xi$, an set the pairwise interactions between
TCR-pMHC bonds at this separation to $\epsilon=-4.5~k_{\rm B}T$,
according to eq.~(\ref{eq:F2}) (which has been derived from
eq.~(\ref{eq:HH})).

In addition to the membrane-mediated interactions between TCR-pMHC
bonds, we also need to calculate the pair PMF between the TCR-pMHC
bonds and proteins that pin the T cell membrane to the actin
cytoskeleton. These interactions are obviously repulsive due to the
large differences in the height of the membrane at the locations of
these two proteins ($h=0$ at the pinning sites compared to
$h=h_0=27\,$nm at the sites of the TCR-pMHC bonds). The repulsive pair
PMF can be derived from the partition function
\begin{equation}
\label{eq:Zb}
Z_B=\int{\cal D}\left[h({\bf r})\right] e^{-\beta{\cal
    H}}\delta\left(h(0)-h_0\right)\delta\left(h(r)\right),
\end{equation}
which differs from eq.~(\ref{eq:Za}) by one height constraint. The
resulting repulsive pair PMF is given by
\begin{equation}
\frac{\Phi_{\rm rep}(r)}{k_{\rm
    B}T}=\frac{1}{2}\left(\frac{h_0}{\Delta}\right)^2\frac{\left(\frac{4}{\pi}\right)^2{\rm
    kei^2}\left(\frac{r}{\xi}\right)}{1-\left(\frac{4}{\pi}\right)^2{\rm
    kei^2}\left(\frac{r}{\xi}\right)}
\label{eq:F2rep}
\end{equation}
and is depicted by the dashed line in Fig.~\ref{fig:figSI2} for
similar values of system parameters. From Fig.~\ref{fig:figSI2} it
is clear that $\Phi_{\rm rep}$ is a purely repulsive potential of
range $r\simeq\xi$ that quickly decays to zero at larger
separations. For $r<\xi$, $\Phi_{\rm rep}$ sharply increases similarly
to an excluded volume potential. We thus conclude that the membrane
curvature itself serves as a source of repulsion between TCR-pMHAC
bonds and pinning proteins, which renders the addition of explicit
excluded volume (hard core) interactions unnecessary in this case. In
the lattice simulations with lattice constant $\xi$, this strong
membrane-mediated repulsion is accounted for by the standard demand
that each lattice site can be occupied by no more than a single
lattice point.


\begin{thebibliography}{99}
\bibitem{footnote1} The utility of this approach was demonstrated in
  {\protect ref.~\cite{Dharan}}, where we employed a more fine-grained
  lattice model with a microscopic spacing of $5\,$nm and an explicit
  representation of the membrane that resulted in TCR-pMHC domains
  with similar densities to those observed in the IS.
\end{thebibliography}
\end{document}